\documentclass[aps,secnumarabic,nobalancelastpage,amsmath,amssymb,
nofootinbib,preprint]{revtex4}
\usepackage{graphics}      % standard graphics specifications
\usepackage{graphicx}      % alternative graphics specifications
\usepackage{longtable}     % helps with long table options
\usepackage{url}           % for on-line citations
\usepackage{bm}            % special 'bold-math' package
\newcommand{\beq}{\begin{equation}}
\newcommand{\eeq}{\end{equation}}
\newcommand{\bea}{\begin{eqnarray}}
\newcommand{\eea}{\end{eqnarray}}

\begin{document}

\title{Quantum Vacuum Fluctuations in a Chromomagnetic-like Background}

\author{V. B. Bezerra\footnote{E-mail:valdir@fisica.ufpb.br}, M. S. Cunha\footnote{E-mail:marcony.cunha@uece.br}, C. R. Muniz\footnote{E-mail:celio.muniz@uece.br}, M. O. Tahim\footnote{E-mail:makarius.tahim@uece.br}}
\affiliation{${}^{*}$Departamento de F\'isica, Universidade Federal da Para\'iba, Cidade Universit\'aria, s/n, Jo\~ao Pessoa-PB, 58051-970, Brazil.\\ ${}^\dagger$Grupo de F\'isica Te\'orica (GFT), Centro de Ci\^encias e Tecnologia, Universidade Estadual do Cear\'a, Av. Dr. Silas Munguba, 1700, 60714-903, Fortaleza-CE, Brazil. \\ ${}^{\ddagger}$\!Universidade Estadual do Cear\'a, Faculdade de Educa\c c\~ao, Ci\^encias e Letras de Iguatu, Rua Deocleciano Lima Verde, s/n Iguatu-CE, Brazil. \\
${}^{\S}$\!Universidade Estadual do Cear\'a, Faculdade de Educa\c c\~ao, Ci\^encias e Letras do Sert\~ao Central, 63900-000, Quixad\'a-CE, Brazil.}

%\ead{marcony.cunha@uece.br}\address{}

%\date{}

\begin{abstract}
In this paper we study the effects associated to quantum vacuum fluctuations of vectorial perturbations of the Abelian SU(2) Yang-Mills field in a static and homogeneous chromomagnetic-like background field, at zero temperature. We use periodic and antiperiodic boundary conditions in order to calculate the Casimir energy by means of the frequency sum technique and of the regularization method based on zeta functions, analyzing its behavior in the weak and strong coupling regimes. We compare the obtained results with the similar ones found for scalar and spinor fields placed in an ordinary magnetic field background. We show that only in the weak coupling regime the non-trivial topology of the system encoded in the antiperiodic boundary conditions changes the nature of the Casimir force with respect to the periodic ones. Considering the weak coupling scenario, we show that the introduction of a third polarization state in the perturbations makes manifest the effects on the Casimir energy due to the coupling with the chromomagnetic-like background field, for both the boundary conditions. Finally, in the strong coupling regime, in which the quantum vacuum is not stable due to the Nielsen-Olesen instabilities, we evaluate the effects of a compact extra dimension on its stabilization.\\\\
\vspace{0.75cm}
\end{abstract}


\maketitle

\section{Introduction}

It is known since the late 1970´s that the so-called Abelian solutions of the Yang-Mills (Y-M) theory are unstable \cite{Mandula,Nielsen,Chang,Sikivie}, {\it i.e.}, there are normal modes for these fields which are associated to the exponential growth of infinitesimal perturbations in the strong coupling regime. Such Abelian solutions represent a gauge choice for the potentials which leaves the field strength unidirectional in the internal space of the group as well as in the ordinary space and can be used to investigate the vacuum instability of Quantum Chromodynamics (QCD) \cite{Savvidy}. These fields keep a formal relationship with electric or magnetic fields, and for this reason they are called chromoelectric or chromomagnetic field. The motivations to study such fields are many, such as the analysis of their behavior in nontrivial topologies \cite{Hansson} as well as of models in which they are seeds for the observed large scale (intergalactic) magnetic fields \cite{Enquist,Bordag,Vadim}.

Nielsen and Olesen \cite{Nielsen} studied the referred instability as well as its effects on the asymptotic freedom and on vacuum polarization, finding the energies of the perturbation modes of the SU(2) Yang-Mills (Y-M) fields in a chromomagnetic-like background. These authors have shown that at least one mode is unstable since the energy eigenvalue becomes imaginary. Other studies indicate that the QCD vacuum is a very complicated system \cite{Nachtmann} (see also references therein), which means that at short distances, below the QCD scale, it presents a nontrivial topological structure, in which may occur states describing a kind of condensed particle, as a superconductor, with energy level below the one associated to states without particles \cite{Chernodub}.

Another phenomenon concerning to the richness and complexity of the vacuum is the Casimir effect, which represents a direct manifestation of the fundamental physical reality named quantum vacuum. It has been exhaustively studied after the prediction of this phenomenon made by H. Casimir \cite{Casimir} as an attractive force that arises between two parallel, infinite, and uncharged metallic plates placed in the perfect vacuum at zero temperature. This force is caused essentially by the shift in the electromagnetic zero point energy due to the presence of the material boundaries. The theoretical and experimental studies related to the phenomenon have grown since then, covering a wide range of boundaries made by a variety of materials and with different geometries, for quantum fields of different spins, and spaces with different metrics and nontrivial topologies \cite{Mostepanenko}.

In particular, the study of the Casimir effect in nontrivial topologies has been made in several contexts from Cosmology \cite{Zhuk,Mostepanenko2,Herondy1,Herondy2,Herondy3}, including universes with compact extra dimensions \cite{Mostepanenko3,Teo}, up to condensed matter physics \cite{Grushin, Rodriguez,Grushin2,Celio}. In the context of alternative theories of gravity, the phenomenon also was studied \cite{Celio2,Celio3,Celio4}, and in General Relativity, by linearizing the field equations and considering real parallel plates. Recently, the Casimir effect was considered as a possible phenomenon which can be used to detect gravitons \cite{Quach}.

The complexity of the QCD vacuum structure has been little explored with relation to the Casimir effect. The works on this theme practically covers only the bag model for the nucleon structure \cite{Milton1,Milton2,Milton3,Chodos,Bardeen}. On the other hand, the possibility of the existence of large scale chromomagnetic fields, at least in the early Universe, demands the investigation of the Casimir effect in such a context, since it also manifests itself as a macroscopic quantum phenomenon.  In the present paper we intend to contribute in this direction, by calculating the vacuum energy of Abelian vectorial perturbations of the SU(2) Y-M field in a homogeneous chromomagnetic-like background field, submitted to periodic and antiperiodic boundary conditions associated to colorless parallel plates. We follow the standard procedure which consists of summing the zero-point energies using the results found in \cite{Nielsen}, and regularizing the obtained energy by the zeta function approach. Then we analyze the behavior of the Casimir energy in both the weak and strong coupling regimes. In the first case, we will show that only the presence of a third polarization mode in the perturbations yields a Casimir energy that depends on both the chromomagnetic-like field and the coupling constant, which nevertheless implies the gauge symmetry breakdown. In the second regime we will show how is the contribution of the stable part of the regularized vacuum energy, and then what are the effects of a compact extra dimension on the stabilization of the Casimir energy as a whole. Comparisons with results found in the literature \cite{Cougo,Erdas,Erdas1} for scalar and spinor fields in the presence of an ordinary magnetic field will be made. Therefore, the present paper seeks to investigate the effects on the boundaries of the quantum vacuum associated to the Abelian SU(2)Yang-Mills field fluctuations in the chromomagnetic-like background, considering also the presence of nontrivial topologies.
The paper is organized as follows: in Section 2, we present the general expression for the regularized vacuum energy and discuss the considered regimes. In section 3, we discuss the results and present some remarks.

\section{Quantum vacuum energy of SU(2) Y-M fluctuations in an uniform chromomagnetic-like field}

In this section, we present the general calculations of the regularized vacuum energy of the vectorial quantum fluctuations, $a_\mu({\bf x},t)$, of the massless Y-M field in the presence of a static and uniform chromomagnetic-like field, $B$. These perturbations are such that $|a_\mu({\bf x},t)|\ll 1$, and thus permitting the linearization of the corresponding equation of motion. This in turn allows us to obtain the eigenfrequencies of the system, which are given by \cite{Nielsen}
\begin{eqnarray}\label{01}
\omega_{kn\epsilon}^2&=&\frac{k^2\pi^2}{a^2}+gB(2n+1-2\epsilon),\\
\omega_{kn\epsilon}^2&=&\frac{4(k+1/2)^2\pi^2}{a^2}+gB(2n+1-2\epsilon),
\end{eqnarray}
for periodic boundary conditions, $a_\mu(x,y,z,t)=a_\mu(x,y,z+a,t)$, and antiperiodic boundary conditions, $a_\mu(x,y,z,t)=-a_\mu(x,y,z+a,t)$, respectively. This latter is quite used in the bag models of nucleons \cite{Chodos,Bardeen}. Here $g$ is the coupling constant, $B$ stands for the chromomagnetic-like background field, $k,n=0,1,2,3...$ and $\epsilon=-1,1$, since the perturbation is initially due to a massless spin one particle. In our model, the colorless plates have area $L^2$ and are separated by a distance $a$. The linearization of the Y-M equations yielding vectorial fluctuations with Abelian symmetry allows us to assume the periodic boundary conditions usually employed in the electromagnetic Casimir effect. On the other hand, we also use the antiperiodic boundary conditions in order to explore in a simple way the non-trivial topology of the vacuum structure and to see the changes in the nature of the Casimir force when compared with the previous case, namely, with periodic conditions taken into account.

Considering initially the periodic boundary conditions, the vacuum energy $\omega^{(0)}$ will be the sum taken over all zero point frequencies, which can be written as
\begin{equation}\label{02}
\omega^{(0)}=\frac{\sqrt{2}(gB)^{3/2}L^2}{4\pi}\sum_{k=1}^{\infty}\sum_{n=0}^{\infty}\left[\sqrt{\frac{k^2\pi^2}{2gBa^2}+\frac{3}{2}+n}+\sqrt{\frac{k^2\pi^2}{2gBa^2}-\frac{1}{2}+n}\right],
\end{equation}
where the degeneracy $gB/2\pi$ per unit area was considered. The modes $k=0$ were excluded, since its contribution to the vacuum energy corresponds to an additive constant, independent of $a$. The expression in Eq. (\ref{02}) is divergent, and we will regularize it using the Hurwitz zeta function $\zeta_H(s,q)=\sum_{n=0}^{\infty}(n+q)^{-s}$, valid for $Re(q)>0$. Thus, we arrive at
\begin{eqnarray}\label{03}
\omega_{reg}^{(0)}&=&\frac{\sqrt{2}(gB)^{3/2}L^2}{4\pi}\lim_{s\rightarrow-1/2}\sum_{k=1}^{\infty}\left[\zeta_H\left(s,\frac{k^2\pi^2}{2gBa^2}+\frac{3}{2}\right)+\zeta_H\left(s,\frac{k^2\pi^2}{2gBa^2}-\frac{1}{2}\right)\right]\nonumber\\
&=&\frac{\sqrt{2}(gB)^{3/2}L^2}{4\pi}\lim_{s\rightarrow-1/2}\sum_{k=1}^{\infty}\left[\zeta_H\left(s,\frac{k^2\pi^2}{2gBa^2}+\frac{1}{2}\right)+\zeta_H\left(s,\frac{k^2\pi^2}{2gBa^2}-\frac{1}{2}\right)\right]+\nonumber\\
&-&\frac{\sqrt{2}(gB)^{3/2}L^2}{4\pi}\sum_{k=1}^{\infty}\sqrt{\frac{k^2\pi^2}{2gBa^2}+\frac{1}{2}}.
\end{eqnarray}

Firstly, we will study the regime that corresponds to the Y-M perturbations in a uniform weak chromomagnetic-like field, considering the weak coupling and short distance between the plates. In such a regime, the stability of the Yang-Mills fields is assured. Thus, taking the following asymptotic expansions of the Hurwitz zeta functions
\begin{eqnarray}\label{04}
 \zeta_{H}(-1/2,x+1/2) &=&-\frac{2}{3} x^{3/2}+\mathcal{O}(x^{-1/2}) \\
 \zeta_{H}(-1/2,x-1/2) &=& -\frac{2}{3} x^{3/2}+\sqrt{x}+\mathcal{O}(x^{-1/2}),
\end{eqnarray}
for $x=k^2\pi^2/g Ba^2 \gg 1$, and substituting the above expressions in Eq. (\ref{03}), observing that $\sqrt{x+1/2}\approx \sqrt{x}$, we get
 \begin{equation}\label{06}
 \omega_{reg}^{(0)}\approx -\frac{\zeta{(-3)}\pi^2L^2}{6a^3},
 \end{equation}
 where $\zeta{(-3)}=1/120$. Notice that we do not have dependence on the chromomagnetic-like field nor on the coupling in the considered regime. In fact, Eq. (\ref{06}) is just the result of the Casimir energy associated to the electromagnetic field.

 In order to see the effects due to both the coupling and the chromomagnetic-like field in this level of approximation, it would be necessary to take into account a third polarization state, namely, the modes with $\epsilon=0$. Thus, let us add to Eq. (\ref{03}) the term $\sum_{k=1}^{\infty}\zeta_H\left(-1/2,x\right)$, so that in the investigated limit we have $\zeta_H\left(-1/2,x\right)\approx -\frac{2}{3} x^{3/2}+\frac{1}{2}\sqrt{x} $, which yields a regularized vacuum energy
 \begin{equation}\label{omega}
 \omega_{reg}^{(0)}\approx -\frac{\pi^2L^2}{480a^3}+\frac{\sqrt{2}\zeta(-1)gBL^2}{8a},
 \end{equation}
 where $\zeta(-1)=-1/12$. It is worth calling attention to the attractive character of the force associated to the correction due to both the coupling constant and the chromomagnetic-like field.

From eigenfrequencies given in Eq. (2), the regularized vacuum energy, when antiperiodic boundary conditions are considered, can be simply obtained by making $a\rightarrow a/2$ and $\zeta(n)\rightarrow \zeta_H(n,1/2)$ in Eq. (\ref{omega}), which results in
 \begin{equation}
 \omega_{reg}^{(0)}\approx \frac{7}{8}\frac{\pi^2L^2}{480a^3}+\frac{\sqrt{2}gBL^2}{192a}.
 \end{equation}
The force has now a repulsive nature, even in the situation in which we do not take into account the third polarization state of the Y-M field perturbations. Thus, through the introduction of antiperiodic boundary conditions we can see how a non-trivial topology affects the regularized vacuum energy in the context of Y-M Abelian fields. The dependence of the regularized vacuum energy on the chromomagnetic-like field in low coupling regime, due to the inclusion of the zero mode, and as a consequence that the Y-M vectorial fluctuations acquire mass, which breaks the gauge invariance of the Yang-Mills fields.

Now, we will consider the opposite regime, {\it i.e.}, when $gB a^2\gg 1$,  in which situation the instabilities of the Y-M fields are manifest. The procedure to regularize the vacuum energy will be something different from that one employed so far. In this case, we consider the identity
 \begin{equation}\label{07}
z^{-s}=\frac{1}{\Gamma(s)}\int_{0}^{\infty}dtt^{s-1}\exp{(-zt)},
\end{equation}
and define
\begin{eqnarray}\label{08}
\tilde{\zeta}_{-1}(s)&\doteq&\frac{1}{\Gamma(s)}\int_{0}^{\infty}dtt^{s-1}\sum_{k=-\infty}^{\infty}\exp{(-k^2\pi^2t/a^2)}\sum_{n=0}^{\infty}\exp{[-gB(2n+3)t]},\\
\tilde{\zeta}_{1}(s)&\doteq&\frac{1}{\Gamma(s)}\int_{0}^{\infty}dtt^{s-1}\sum_{k=-\infty}^{\infty}\exp{(-k^2\pi^2t/a^2)}\sum_{n=0}^{\infty}\exp{[-gB(2n-1)t]},
\end{eqnarray}
 for periodic boundary conditions, where the sum over $k$ means that it is taken over all the integers, in such a way that a factor $1/2$ must be considered. The subindexes are related to the spin $\epsilon$. Thus, we have
\begin{equation}
\omega_{reg}^{(0)}=\frac{(gB)L^2}{8\pi}\lim_{s\rightarrow-1/2}[\tilde{\zeta}_{-1}(s)+\tilde{\zeta}_{1}(s)].
\end{equation}
Making the change of variable $t\rightarrow (\sqrt{gB}/a)t$ and using the Poisson resummation formula
\begin{equation}\label{09}
\sum_{m=-\infty}^{\infty}\exp{(-m^2rt)}=\sum_{m=-\infty}^{\infty}(\pi/rt)^{1/2}\exp{(-m^2\pi^2/rt)},
\end{equation}
Eq. (\ref{08}) turns into
\begin{equation}\label{10}
\zeta_{-1}(s)=\frac{(gB)^{(1-2s)/4}a^{s+1/2}}{\Gamma(s)\sqrt{\pi}}\int_{0}^{\infty}dtt^{s-3/2}\sum_{k=-\infty}^{\infty}\exp{(-k^2\sqrt{gB}a/t)}\sum_{n=0}^{\infty}\exp{[-(2n+3)\sqrt{gB}at]}.
\end{equation}
The last sum in the integral given by Eq. (15) is equal to $e^{-\sqrt{gB}a t}/(e^{2 \sqrt{gB}a t}-1)$, which is reduced to $e^{-3\sqrt{gB}at}$ in the considered approximation. Taking into account the leading terms of the first sum in Eq. (\ref{10}), {\it i.e.}, those ones for which $k=\pm 1$, we can solve the remaining integral and obtain the following result
 \begin{equation} \label{15}
 \lim_{s\rightarrow -1/2}\tilde{\zeta}_{-1}(s)=-\frac{2(g B)^{1/2}\sqrt{3}}{\pi}K_{1}(2\sqrt{3gB}a),
 \end{equation}
where $K_1(z)$ is the modified Bessel function of first kind.

Regarding $\tilde{\zeta}_{1}(s)$, we arrive at a non-convergent integral, since it contains an element of instability in the Y-M fields - recall that the eigenfrequencies with $n=0$ and $\epsilon=1$ become imaginary in the regime under consideration. Thus, in order to evaluate the stable part of the Casimir energy, we must unconsider the term $n=0$ from $\sum_{n=0}^{\infty} \exp{[-(2n-1)\sqrt{gB}at]}\approx \exp{(\sqrt{gB}at)}$. In this case, $\tilde{\zeta}_1(s)$ vanishes, eliminating the divergent part of the vacuum energy in the cited regime. The stable part of the regularized vacuum energy is, therefore
 \begin{equation} \label{17}
 \omega^{(0)}_{reg,st}\approx \frac{g B L^2}{8\pi}\zeta_{-1}(-1/2)=-\frac{\sqrt{3}(gB)^{3/2}L^2}{4\pi^2}K_1(2\sqrt{3gB}a).
 \end{equation}
This stable contribution to the vacuum energy is associated to an attractive force, which is very weak, going to zero in the limit of very strong fields, which means that the Olesen-Nielsen instability is absolutely dominant in that regime, as expected.  With respect to the antiperiodic boundary conditions, the results above are the same, except for numerical factors, and the signal of the force does not change. The unstable part remains for $\epsilon=1$. It is worth noticing that the expression for the regularized vacuum is slightly different for a charged and massless scalar particle in a magnetic field, including just also others numerical factors.

In order to overcome the instability of the Casimir energy in that regime, it would be necessary to evaluate the influence of the mass of the fluctuations on the system, adding $m^2$ to Eq.(1), which will yield a factor $\exp{(-m^2t)}$ in the integrals of Eqs. (11) and (12). Then, this latter will converge provided $m\geq \sqrt{g B}$, which means that a large value for the mass contributes to stabilize of the Casimir energy, in the regime under consideration. We must also consider the spin term $\epsilon=0$ due to this mass, and a third zeta function given by
\begin{equation}
\tilde{\zeta}_{0}(s)\doteq\frac{1}{\Gamma(s)}\int_{0}^{\infty}dtt^{s-1}\sum_{k=-\infty}^{\infty}\exp{[-(m^2+k^2\pi^2)t/a^2]}\sum_{n=0}^{\infty}\exp{[-gB(2n+1)t]},
\end{equation}
must be added to the terms in the brackets of Eq. (13). Thus, after adopting the same procedure above, we find the Casimir energy
\begin{eqnarray}
 \omega^{(0)}_{reg}&\approx &-\frac{(gB)L^2}{4\pi^2}[\sqrt{3gB+m^2}K_1(2\sqrt{3gB+m^2}a)+\sqrt{m^2-gB}K_1(2\sqrt{m^2-gB}a)\nonumber\\
 &+&\sqrt{gB+m^2)}K_1(2\sqrt{gB+m^2}a)].
\end{eqnarray}
With respect to the antiperiodic boundary conditions, the modifications occur only in the numerical factors, and thus the analytical results are the same.
\begin{figure}[!h]
\centering
\includegraphics[scale=0.55]{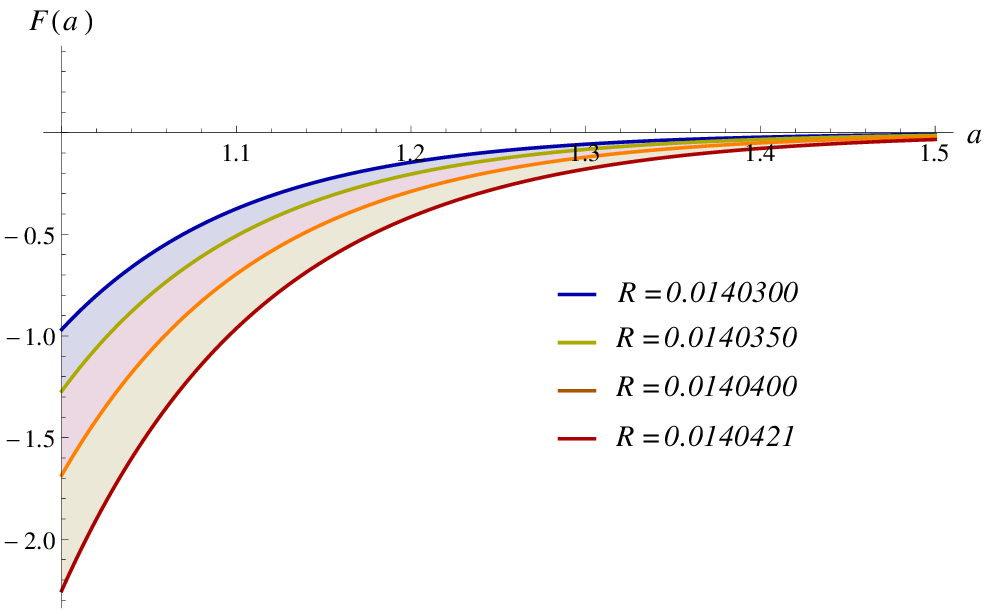}
\caption{From top to bottom, plots of the Casimir force per area unit as a function of the distance between the plates, considering increasing values for the extra dimension radius $R$, respectively. In all these plots, we have used $gB=5\times 10^3$.}
 \label{umg}
\end{figure}

Taking into account the aforementioned mass breaks the gauge invariance of the Yang-Mills field, and a natural mechanism to generate it was proposed in \cite{Chivukula}, which involves a compact extra dimension, without the necessity of introducing Higgs fields. \cite{Chivukula}. The idea is to consider the existence of an extra dimension {\it \`a la} Kaluza-Klein in order to obtain a compactified five-dimensional massless Yang-Mills theory, as seen in the usual four dimensions, which would be thus linearized according to the Nielsen-Olesen model. The corresponding dimensional reduction would yield a set of adjoint vector bosons with mass $m_{\ell} = \ell/R$, where $\ell=0,1,2...$. For convenience, we consider compactifying the fifth dimension to a line segment $0\leq x^5 \leq \pi R$, which can be done consistently by an orbifold projection \cite{Chivukula}. As the eigenfrequencies (\ref{01}) we should add the term $1/R^2$, corresponding to the first non-zero mode of the Y-M field in the Kaluza-Klein approach. With this procedure, we arrive at
\begin{eqnarray}
\omega^{(0)}_{reg}&\approx&-\frac{(g B)L^2}{4\pi^2 R}\left[\sqrt{1+3gBR^2}K_1\left(\frac{2a}{R}\sqrt{1+3gBR^2}\right)
+\sqrt{1-gBR^2}K_1\left(\frac{2a}{R}\sqrt{1-gBR^2}\right)\right]\nonumber\\
&-&\frac{(g B)L^2}{4\pi^2 R}\left[\sqrt{1+gBR^2}K_1\left(\frac{2a}{R}\sqrt{1+gBR^2}\right)\right].
\end{eqnarray}
Thus, the compact extra dimension would stabilize the regularized quantum vacuum of the Y-M fluctuations below a critical radius, given by $R_c= (gB)^{-1/2}$. In the Fig.1, we depict the Casimir force $F(a)=-\partial\omega_{reg}^{(0)}/\partial a$ as a function of the distance between the plates, for certain values of the extra dimension radius, which are below the value of the critical radius, defined from the fixed quantity $gB=5\times 10^3$. Notice that the attractive force increases when the radius increases.

 \section{conclusions and remarks}

We have computed the regularized quantum vacuum energy of SU(2) Y-M vectorial perturbations in a homogeneous and static chromomagnetic-like field background, using periodic and antiperiodic boundary conditions, at zero temperature. We analyzed its behavior in both the regimes of weak and strong coupling, using different techniques based on the zeta function regularization.

In the first case, the system is free of the Nielsen-Olesen instabilities. Initially, by taking into account periodic boundary conditions we analyzed the Y-M vectorial perturbations with two degrees of freedom, obtaining a Casimir energy that is trivially identical to that one of a pure electromagnetic field. Thus, in order to show up the effects of a chromomagnetic-like field in the weak approximation, it was necessary to take into account effects of the spin by introducing a third polarization state, which resulted in a correction to the Casimir energy proportional to $a^{-1}$, amplifying the attractive force previously found, despite this intensification to be due to the gauge symmetry breakdown. Therefore, if we intend to keep the gauge invariance of the system in the low coupling regime, the Casimir energy must be free of the background field effects, at least in the first order approximation. On the other hand, we found that, in this regime, and solely in it, a nontrivial topological structure of the quantum vacuum modelled by antiperiodic boundary conditions applied on the Y-M vectorial perturbations changes the signal of the Casimir force, as well as some numerical factors, but preserving its dependence on the separation between the colorless plates.

We have also investigated the strong coupling regime, in which the Nielsen-Olesen instabilities are unavoidable. As expected, we found a divergent part in the Casimir energy and a contribution of the stable part expressed by Eq. (\ref{17}), and can be once more compared with the cases of a massless charged scalar particle or a spinorial one under the influence of an ordinary constant magnetic field in this regime \cite{Erdas}. In this case, the Casimir energies differ from those ones of the Y-M perturbations in a chromomagnetic-like field just by numerical factors. The antiperiodic boundary conditions yield the same result, except by numerical factors, too. In this regime, a nontrivial topology introduced by means of these boundary conditions does not change the character of the Casimir force associated to the vacuum energy stable part.

We have shown that a very large mass associated to the fluctuations stabilizes the quantum vacuum energy, which is equivalent to consider the simplest form of a compact extra dimension, which means that the space now owns a non-trivial orbifold topology with radius $R$. In this case, the stabilization of the system occurs provided that $R\leqslant 1/\sqrt{gB}$, and thus we conclude that, at zero temperature, a possible mechanism to stabilize the quantum vacuum of Y-M perturbations would be realized through the existence of a compact extra dimension. Such a stabilization of SU(2) Y-M vacuum energy was already studied in the context of other non-trivial topologies, as the one of the torus \cite{Hansson}. As future perspective for the present investigation, we propose to study the effects of more general compact extra dimensions as well as of the temperature on the mentioned stabilization.

 \section*{Acknoledgements} The authors would like to thank CNPq by partial support.

\end{document}